\documentclass{jpsj2}

\usepackage{graphicx}

\title{Perturbation Theory of High-$T_{\rm c}$ Superconductivity 
in Iron Pnictides}

\author{Takuji \textsc{Nomura}\thanks{E-mail address: nomurat@spring8.or.jp}}

\inst{
Synchrotron Radiation Research Center, 
Japan Atomic Energy Agency, Sayo, Hyogo 679-5148\\
JST, TRIP, Sanbancho, Chiyoda, Tokyo 102-0075
}

\recdate{\today}

\abst{
The high-transition-temperature (high-$T_{\rm c}$) 
superconductivity discovered recently 
in iron pnictides is analyzed within a perturbation theory. 
Specifically, the probable pairing symmetry, 
the doping dependence of the transition temperature 
and the pairing mechanism are studied by solving the Eliashberg equation 
for multi-band (2- and 5-band) Hubbard models 
with realistic electronic structures. 
The effective pairing interaction is expanded perturbatively 
in the on-site Coulomb integrals up to third order. 
Our perturbative weak-coupling approach shows that 
sufficiently large eigenvalues of the Eliashberg equation are 
obtained to explain the actual high transition temperatures 
by taking realistic on-site Coulomb integrals in the 5-band model. 
Thus, unconventional (non-phonon-mediated) 
superconductivity is highly likely to be realized. 
The superconducting order parameter does not change 
its sign on the Fermi surfaces, but it does change 
between the electron and hole Fermi surfaces. 
Consequently, the probable pairing symmetry 
is always ``a nodeless extended $s$-wave symmetry
(more specifically, an $s_\pm$-wave symmetry)'' 
over the whole parameter region that we investigated. 
It is suggested that the 2-band model is insufficient 
to explain the high values of $T_{\rm c}$.}

\kword{
iron-pnictide superconductors, superconducting mechanism, 
pairing symmetry, perturbation theory, multi-band Hubbard model}

\begin{document}
\maketitle

\section{Introduction}
\label{sc:Introduction}

The recent discovery of high-$T_{\rm c}$ superconductivity 
in iron pnictides and related compounds has strongly intensified 
the research activity in solid-state physics. 
The highest transition temperature attained in iron pnictides is 
about 56 K to date, which is currently the highest reported 
superconducting transition temperature among noncuprate superconductors. 

A prototypical iron-pnictide superconductor LaFeAsO$_{1-x}$F$_x$ 
($x=$0.05-0.12, $T_{\rm c}=26$ K) was discovered 
by Kamihara et al~\cite{ref:KamiharaY2008}. 
Before the discovery of the arsenic superconducting system 
LaFeAsO$_{1-x}$F$_x$, it had been discovered 
that LaFePO$_{1-x}$F$_x$, where As in the arsenic system is 
replaced by P, is also a superconductor 
with $T_{\rm c}=$3.2-6 K~\cite{ref:KamiharaY2006}. 
The LaFeAsO$_{1-x}$F$_x$ crystals are formed 
by stacking LaOF and FeAs layers alternately. 
Each iron atom is surrounded tetrahedrally by four As atoms, 
and each FeAs layer is constructed by connecting the edges 
of the As tetrahedra. In the undoped case $x=0$, 
LaFeAsO appears to be a poor metal, and undergoes 
a structural transition from tetragonal to orthorhombic 
at 155 K and spin-density-wave (SDW) ordering below 137 K 
(with a wave vector ${\mib q}=(\pi, \pi, \pi)$ and 
a magnetic moment about 0.3-0.4$\mu_B$ 
on each iron atom at 0 K)~\cite{ref:DeLaCruzC2008}. 
By substituting oxygen with fluorine, 
doped electrons cause metallic conduction with the disappearance 
of both the SDW and structural transitions for $x>0.05$, 
and superconductivity occurs below about 26 K~\cite{ref:KamiharaY2008}. 
Moreover, subsequent studies clarified that replacement 
of La by other rare earth elements increases 
the transition temperature markedly: 
41 K for Ce~\cite{ref:ChenGF2008}, 
52 K for Pr~\cite{ref:RenZA2008a}, 
52 K for Nd~\cite{ref:RenZA2008b}, and 
55 K for Sm~\cite{ref:ChenXH2008, ref:RenZA2008c}. 
Superconductivity has also been discovered  
in fluorine-free systems, including {\sl R}FeAsO$_{1-y}$ 
({\sl R}=La, Ce, Pr, Nd, Sm, Gd. 
$T_{\rm c}=31$-55 K)~\cite{ref:KitoH2008, 
ref:RenZA2008d, ref:LeeCH2008, ref:YangJ2008}, 
Gd$_{1-x}$Th$_x$FeAsO ($T_{\rm c}=$56 K)~\cite{ref:WangC2008}. 
Hole-doped La$_{1-x}$Sr$_x$FeAsO also shows 
superconductivity with $T_{\rm c}=$25 K~\cite{ref:WenHH2008}. 
Thus, a superconducting phase exists not only in the electron-doped 
side but also in the hole-doped side, and it appears that the superconducting 
transition is inhibited due to the structural distortion and SDW ordering 
only near the undoped case $x=0$. 
It is natural to expect that superconductivity 
in all of these compounds occurs basically by the same mechanism, 
in spite of the quantitative difference in $T_{\rm c}$. 

In addition to {\sl R}FeAsO$_{1-x}$F$_x$ ({\sl R}: rare-earth element) series, 
many other related superconducting iron compounds without oxygen 
have been discovered, e.g., 
Ba$_{1-x}$K$_x$Fe$_2$As$_2$ ($T_{\rm c}=$38 K)~\cite{ref:RotterM2008}, 
Li$_{1-x}$FeAs ($T_{\rm c}=18$ K)~\cite{ref:TappJH2008, ref:WangXC2008}, 
and Ca$_{1-x}$Na$_x$Fe$_2$As$_2$ ($T_{\rm c}=$20 K)~\cite{ref:WuG2008}. 
In these related compounds, FeAs layers are separated not 
by layers of rare-earth elements and oxygen, but by layers 
of BaK, Li, or other various elements. 
Therefore, we naturally consider that the feature essential 
for the superconductivity is the existence of FeAs layers, 
while the remaining elements may play only a relatively minor 
role of a carrier reservoir or spacing between the FeAs layers. 
In fact, similar superconductivity has also been observed 
in $\alpha$-FeSe ($T_{\rm c}=$8 K)~\cite{ref:HsuFC2008}, 
in which the valence number of the iron atoms may 
be the same as that in iron pnictides but no spacing elements 
are inserted between the FeSe layers. 
This fact suggests that only the layers of square networks constructed 
from iron atoms may be necessary for superconductivity. 

Electronic structure calculations for some 
of the above-mentioned iron-based superconductors 
have already been carried out in many 
studies~\cite{ref:LebegueS2007, ref:SinghDJ2008, 
ref:BoeriL2008, ref:KurokiK2008, ref:IshibashiS2008, 
ref:NekrasovIA2008, ref:MaF2008}. 
According to these studies, the electronic states near the Fermi level 
are formed dominantly by the Fe3d electronic states. 
Thus, iron pnictides may be considered to be typical strongly 
correlated electron systems, in which the strong Coulomb interaction 
among the iron 3d electrons can affect the electronic properties significantly. 
The Fermi surface consists of hole pockets around the $\Gamma$ point 
and electron pockets around the M point, qualitatively 
consistent with the results of angle-resolved photoemission (ARPES) 
experiments\cite{ref:DingH2008, ref:KondoT2008, ref:LiuC2008} 
and more recent quantum oscillation experiments\cite{ref:ColdeaAI2008}. 

An important step toward understanding this newly discovered 
iron-pnictide superconductivity will be to clarify 
whether it is conventional phonon-mediated superconductivity 
or unconventional superconductivity such as that in copper oxides. 
There are already several reasons why unconventional 
pairing may be realized in iron pnictides: 
(i) $T_{\rm c}$ is very high, compared 
with conventional phonon-mediated BCS superconductors, 
(ii) Fe3d states dominate the majority of the density 
of states near the Fermi level~\cite{ref:SinghDJ2008, ref:BoeriL2008}, 
as mentioned above, and therefore the electron correlation originating 
from the Coulomb interaction on the relatively localized Fe3d bands 
is considered to be strong in these systems, 
and (iii) electron-phonon coupling is expected to be weak 
according to first-principles calculations~\cite{ref:BoeriL2008}. 
Thus, it will be a very interesting issue 
to determine whether or not we can understand iron-pnictide 
superconductivity as a natural result of electron correlations, 
as in the case of unconventional superconductivity 
in many other strongly correlated systems. 

So many theoretical works on iron-pnictide superconductivity 
have already been presented that we cannot review all of them here. 
Many of them focus on the pairing mechanism and pairing symmetry. 
Among the proposed pairing symmetries, the most promising pairing 
symmetry is a spin-singlet extended $s$-wave ($s_\pm$-wave) symmetry, 
in which the superconducting order parameter changes its sign 
between the Fermi surfaces. 
This pairing symmetry has been suggested in considerations 
based on the magnetic susceptibility~\cite{ref:MazinII2008}, 
random-phase approximation (RPA)~\cite{ref:KurokiK2008, 
ref:YanagiY2008}, renormalization group (RG) 
calculations~\cite{ref:WangF2008, ref:ChubukovAV2008}, 
fluctuation-exchange approximation (FLEX)~\cite{ref:IkedaH2008} 
and other studies~\cite{ref:SeoK2008}. 
The $s_\pm$-wave pairing symmetry appears now to be consistent 
with the results of most experiments, 
including those based on ARPES~\cite{ref:DingH2008, ref:KondoT2008}, 
penetration depth~\cite{ref:HashimotoK2008}, 
and neutron scattering~\cite{ref:ChristiansonAD2008}. 
A recent theoretical study suggests that the sign reversal 
of the $s_\pm$-wave pairing symmetry 
is consistent also with nonmagnetic impurity effects~\cite{ref:SengaY2008}. 
Further investigations are still desirable to confirm that this pairing symmetry 
is indeed realized in iron-pnictide superconductors. 
Our present theoretical study will provide additional support 
for this pairing symmetry. 

In the present work we present a microscopic theory 
of iron-pnictide superconductivity. 
We take a weak-coupling approach on the basis 
of third-order perturbation theory with respect to the electron Coulomb interaction 
among the Fe3d electrons. 
This perturbative approach has been applied to many realistic 
electronic structures in order to discuss the pairing symmetry 
and pairing mechanism in real strongly correlated electron systems, 
and has succeeded in explaining various exotic pairing states, 
including the $d_{x^2-y^2}$-wave states in cuprate 
and organic superconductors~\cite{ref:HottaT1994,ref:JujoT1999,ref:NomuraT2001}, 
the triplet $p$-wave state in Sr$_2$RuO$_4$~\cite{ref:NomuraT2000,ref:NomuraT2002}, 
the $d_{x^2-y^2}$-wave states in several heavy-fermion 
superconductors~\cite{ref:IkedaH2002, ref:NisikawaY2002, ref:YanaseY2003}. 
Therefore, it is interesting to also apply this approach 
to iron-pnictide superconductors. 
Specifically, the probable pairing symmetry, 
the momentum and band dependences of the superconducting order parameter, 
the doping dependence of the transition temperature, 
and the pairing mechanism are investigated by solving 
the Eliashberg equation for 2- and 5-band Hubbard models 
with realistic electronic structures. 
The effective pairing interaction is expanded perturbatively 
in the on-site Coulomb integrals up to third order. 
Our weak-coupling approach shows that sufficiently high values 
of $T_{\rm c}$ are obtained for realistic values 
of the on-site Coulomb interaction in the 5-band models 
in order to explain the experimentally observed high values of $T_{\rm c}$. 
Thus, it is concluded that unconventional (non-phonon-mediated) 
pairing is highly likely to be realized 
in iron-pnictide superconductivity. 
The superconducting order parameter does not change 
its sign on the Fermi surfaces without any nodes, 
but it does change between the electron and hole Fermi pockets. 
Consequently, the probable pairing symmetry is a spin-singlet 
extended $s$-wave or, more specifically, an $s_\pm$-wave symmetry. 
The calculated transition temperatures do not depend 
so sensitively on doping concentration, 
consistent with the results of experiments, 
but should drop suddenly around doping levels 
where some electron or hole pockets disappear. 
This suggests that nesting between the electron 
and hole Fermi pockets is responsible 
for this unconventional pairing. 
It is suggested that the 2-band model is insufficient 
to explain the high transition temperatures. 
The present article is a detailed version 
of our previous short article~\cite{ref:NomuraT2008}. 

\section{Formulation}
\label{sc:Formulation}

In the present work, we use multi-band tight-binding models 
to describe the electronic structure near the Fermi level. 
Taking account of the two-dimensionality of the actual crystal 
and electronic structure, we use two-dimensional tight-binding models. 
The unit cell in a single FeAs layer contains two iron atoms. 
Here we should note that the electronic structure near the Fermi level 
is dominated by the Fe3d orbitals, as many electronic structure 
calculations demonstrate, and will be reproduced effectively 
by taking only the Fe3d-like local orbitals 
(or more specifically, maximally localized Wannier orbitals, 
as presented in ref.~\ref{ref:NakamuraK2008}). 
As Kuroki et al. pointed out~\cite{ref:KurokiK2008}, 
by extracting only Fe3d-like states, 
we can effectively construct a tight-binding model 
on a simple two-dimensional square lattice consisting of only iron sites. 
This simplified scheme helps us greatly by reducing 
the number of energy bands to be considered. 
Thus, the noninteracting part of the Hamiltonian is expressed 
as follows: 
\begin{equation}
H_0=\sum_{i,j}\sum_{\ell,\ell'}\sum_{\sigma} 
t_{i\ell, j\ell'} c_{i\ell\sigma}^{\dag} c_{j\ell'\sigma},
\end{equation}
where $i$ and $j$ denote iron sites 
on the square lattice, $\ell (\ell')$ 
denote Fe3d-like localized orbitals, 
and $c_{i\ell\sigma}$ ($c_{i\ell\sigma}^{\dag}$) 
is the annihilation (creation) operator of electrons 
of Fe3d-like orbital $\ell$ at site $i$ with spin $\sigma$. 
Here note again that the index specifying each of the two iron 
atoms in a unit cell is not necessary. 
Specific choices of $t_{i\ell, j\ell'}$ are given 
in \S\ref{sc:5-band} and \S\ref{sc:2-band}. 

We can diagonalize $H_0$ by using a unitary transformation 
in the form 
\begin{equation}
H_0=\sum_{{\mib k}} \sum_a \sum_{\sigma} 
E_a({\mib k}) c_{{\mib k} a \sigma}^{\dag} c_{{\mib k} a \sigma}, 
\end{equation}
where the index $a$ specifies each of the diagonalized bands 
and $\mib{k}$ is momentum. 
$E_a(\mib{k})$ is the diagonalized-band energy, 
and $c_{{\mib k} a \sigma}$ is related to $c_{i\ell\sigma}$ by 
\begin{equation}
c_{i\ell\sigma} = \frac{1}{\sqrt{N}} 
\sum_{{\mib k}} \sum_a U_{\ell a}(\mib{k}) 
e^{{\rm i}{\mib k}\cdot{\mib r}_i} c_{{\mib k} a \sigma}, 
\label{eq:H_0}
\end{equation}
where $N$ is the number of iron lattice sites 
and $U_{\ell a}(\mib{k})$ represents the elements 
of the diagonalization matrix. 

The electrons on the local orbitals 
will be affected by the on-site Coulomb interaction. 
The interacting part is given in the following form: 
\begin{eqnarray}
H'&=& \frac{U}{2}\sum_{i}\sum_{\ell}\sum_{\sigma \neq \sigma'}
c_{i\ell\sigma}^{\dag} c_{i\ell\sigma'}^{\dag} 
c_{i\ell\sigma'} c_{i\ell\sigma} 
+ \frac{U'}{2}\sum_{i}\sum_{\ell \neq \ell'}\sum_{\sigma, \sigma'} 
c_{i\ell\sigma}^{\dag} c_{i\ell'\sigma'}^{\dag} 
c_{i\ell'\sigma'} c_{i\ell\sigma} \nonumber\\
&& + \frac{J}{2}\sum_{i}\sum_{\ell \neq \ell'}\sum_{\sigma, \sigma'} 
c_{i\ell\sigma}^{\dag} c_{i\ell'\sigma'}^{\dag} 
c_{i\ell\sigma'} c_{i\ell'\sigma} 
+ \frac{J'}{2}\sum_{i}\sum_{\ell \neq \ell'}\sum_{\sigma \neq \sigma'} 
c_{i\ell\sigma}^{\dag} c_{i\ell\sigma'}^{\dag} 
c_{i\ell'\sigma'} c_{i\ell'\sigma}, 
\label{eq:H'}
\end{eqnarray}
where $U$, $U'$, $J$, and $J'$ are the Coulomb integrals 
for the intraorbital Coulomb interaction, 
the interorbital Coulomb interaction, 
the Hund's coupling, and the interorbital pair transfers, respectively. 
The total Hamiltonian is given by
\begin{equation}
H=H_0+H'.
\end{equation}

Propagation of the electrons of band $a$ in the normal state 
is described by the normal Green's function: 
\begin{equation}
G_a^{(0)}(k)=\frac{1}{{\rm i} \omega_n - E_a(\mib{k})}, 
\label{eq:Ga}
\end{equation}
where $k=(\mib{k}, {\rm i} \omega_n)$ and 
$\omega_n$ is the fermionic Matsubara frequency: 
$\omega_n= (2n+1) \pi T$. 
Superconducting transition points can be determined 
by the linearized Eliashberg equation: 
\begin{equation}
\Delta_{a\sigma_1\sigma_2}(k) = 
- \frac{T}{N} \sum_{k'}\sum_{a'}\sum_{\sigma_3\sigma_4} 
V_{a\sigma_1\sigma_2, a'\sigma_4\sigma_3}(k, k') 
|G_{a'}^{(0)}(k')|^2 \Delta_{a'\sigma_3\sigma_4}(k'), 
\label{eq:Eliashberg}
\end{equation}
where $\sigma_i$ are spin indices, $T$ is temperature, 
$\Delta_{a\sigma_1\sigma_2}(k)$ is the anomalous self-energy 
on band $a$, and $V_{a\sigma_1\sigma_2, a'\sigma_3\sigma_4}(k, k')$ 
is the effective pairing interaction. 
We have used the shorthand notation 
$\sum_{k}=\sum_{\mib k}\sum_{\omega_n}$.
At superconducting transition points, 
the Eliashberg equation has a non-self-evident solution, 
i.e., $\Delta_{a\sigma\sigma'}(k) \not\equiv 0$. 
Replacing the left-hand side of eq.~(\ref{eq:Eliashberg}) 
by $\lambda \Delta_{a\sigma_1\sigma_2}(k)$, 
then eq.~(\ref{eq:Eliashberg}) is regarded 
as an eigenvalue equation with the eigenvalue $\lambda$ 
and eigenfunction $\Delta_{a\sigma_1\sigma_2}(k)$, 
and transition points are determined 
by the condition $\lambda_{\rm max} = 1$, 
where $\lambda_{\rm max}$ is the maximum eigenvalue. 
Since $\lambda_{\rm max}$ is usually a monotonically 
decreasing function of $T$, 
therefore the superconducting instability is already attained 
for $\lambda_{\rm max}>1$ and normal state is still favored 
for $\lambda_{\rm max}<1$. 
The momentum dependence and spin symmetry of the eigenfunction 
$\Delta_{a\sigma_1\sigma_2}(k)$ giving 
$\lambda_{\rm max}$ characterize the most favorable pairing 
state. 

\begin{figure}
\begin{center}
\includegraphics[width=0.8\linewidth]{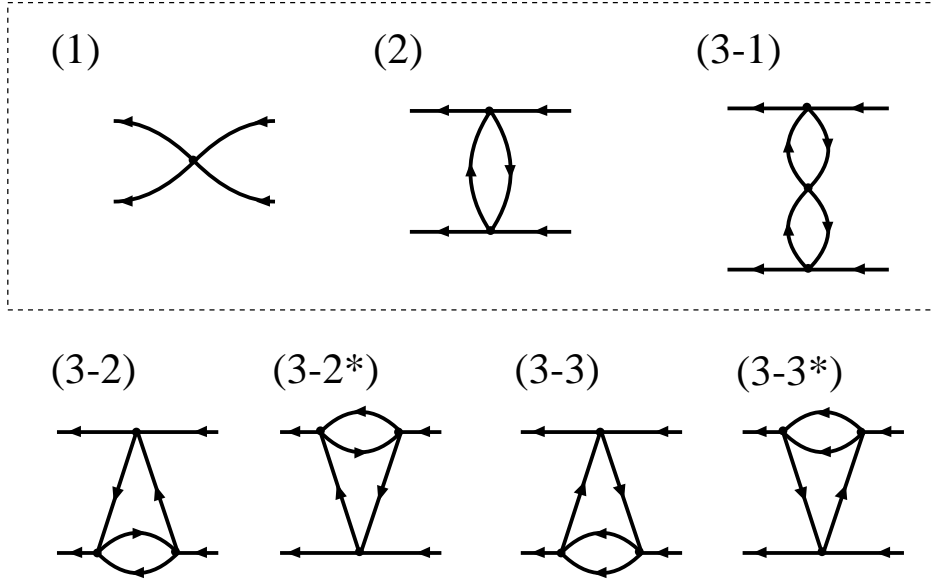}
\end{center}
\caption{
Diagrammatic representation of the effective pairing interaction 
$V_{\zeta_1\zeta_2, \zeta_3\zeta_4}(k, k')$ 
in the third-order perturbation theory. 
Each vertex and solid line with an arrowhead 
represent the on-site Coulomb interaction 
$\Gamma_{\zeta_1\zeta_2,\zeta_3\zeta_4}^{(0)}$ 
and the Green's function $G_{\zeta_1\zeta_2}^{(0)}(k)$, respectively. 
The contributions enclosed by the broken rectangle, 
(1), (2), and (3-1), are also included in RPA calculations, 
while the remaining contributions, (3-2), (3-2$^*$), 
(3-3), and (3-3$^*$) are the vertex corrections 
and are not included within the RPA. 
}
\label{graphs}
\end{figure}

Evaluation of the effective pairing interaction 
$V_{a\sigma_1\sigma_2, a'\sigma_3\sigma_4}(k, k')$ 
is a quantum many-body problem. 
The pairing interaction is expressed in the orbital 
representation by using the diagonalization matrix elements 
$U_{\ell a}({\mib k})$ as follows: 
\begin{eqnarray}
V_{a\sigma_1\sigma_2, a'\sigma_3\sigma_4}(k, k') 
= \sum_{\{\ell_n\}} 
U_{\ell_1 a}^*({\mib k}) U_{\ell_2 a}^*(-{\mib k}) 
V_{\zeta_1\zeta_2, \zeta_3\zeta_4}(k, k')
U_{\ell_3 a'}(-{\mib k}') U_{\ell_4 a'}({\mib k}'), 
\label{eq:Vaa}
\end{eqnarray}
where $\zeta_n$ is the orbital-spin combined index, 
i.e., $\zeta_n=(\ell_n, \sigma_n)$, 
$U^*_{\ell a}({\mib k})$ is the complex conjugate 
of $U_{\ell a}({\mib k})$, and 
the summation with $\{\ell_n\}$ denotes 
that with respect to all of $\ell_1$, $\ell_2$, $\ell_3$, and $\ell_4$. 
Here we define the Green's function in the orbital representation by 
\begin{equation}
G_{\zeta_1\zeta_2}^{(0)}(k)=\sum_a 
U_{\ell_1 a}({\mib k})U_{\ell_2 a}^*({\mib k})
G_a^{(0)}(k) \delta_{\sigma_1\sigma_2}. 
\label{eq:Gll}
\end{equation}
We note that $H'$ can be expressed in the form 
\begin{equation}
H' = \frac{1}{2} \sum_i \sum_{\{\zeta_n\}} 
I_{\zeta_1\zeta_2,\zeta_3\zeta_4} 
c_{i\ell_1\sigma_1}^{\dag} c_{i\ell_2\sigma_2}^{\dag} 
c_{i\ell_3\sigma_3} c_{i\ell_4\sigma_4}, 
\end{equation} 
with 
\begin{equation}
\left.
\begin{array}{lll}
I_{(\ell\sigma)(\ell\bar{\sigma}),(\ell\bar{\sigma})(\ell\sigma)} &= &U\\
I_{(\ell\sigma)(\ell'\sigma),(\ell'\sigma)(\ell\sigma)} &= &U'-J\\
I_{(\ell\sigma)(\ell'\bar{\sigma}),(\ell'\bar{\sigma})(\ell\sigma)} &= &U'\\
I_{(\ell\sigma)(\ell'\bar{\sigma}),(\ell\bar{\sigma})(\ell'\sigma)} &= &J\\
I_{(\ell\sigma)(\ell\bar{\sigma}),(\ell'\bar{\sigma})(\ell'\sigma)} &= &J'
\end{array}\right\}, 
\end{equation}
and the other elements are zero, 
where $\ell\neq\ell'$ and $\sigma\neq\bar{\sigma}$. 
The summation with $\{\zeta_n\}$ denotes that with respect 
to all of the orbital and spin indices, 
$\ell_n$ and $\sigma_n$. 
It is useful to define the antisymmetrized bare Coulomb vertex by 
\begin{equation}
\Gamma_{\zeta_1\zeta_2,\zeta_3\zeta_4}^{(0)}=
I_{\zeta_1\zeta_2,\zeta_3\zeta_4}-I_{\zeta_1\zeta_2,\zeta_4\zeta_3}. 
\end{equation}
We expand $V_{\zeta_1\zeta_2, \zeta_3\zeta_4}(k, k')$ 
perturbatively to third order in $H'$, 
using $G_{\zeta_1\zeta_2}^{(0)}(k)$ and 
$\Gamma_{\zeta_1\zeta_2,\zeta_3\zeta_4}^{(0)}$. 
Diagrammatic representations of the perturbation terms 
are displayed in Fig.~\ref{graphs}. 
The Eliashberg equation in the orbital representation 
is written using the anomalous 
Green's function $F_{\zeta_1\zeta_2}(k)$ as 
\begin{equation}
\Delta_{\zeta_1\zeta_2}(k) = - \frac{T}{N}
\sum_{k'}\sum_{\zeta_3\zeta_4}
F_{\zeta_3\zeta_4}(k')
V_{\zeta_1\zeta_2, \zeta_4\zeta_3}(k, k'). 
\label{eq:Eliashberg2}
\end{equation}
Through the expansion of $V_{\zeta_1\zeta_2, \zeta_4\zeta_3}(k, k')$, 
the anomalous self-energy $\Delta_{\zeta_1\zeta_2}(k)$ 
is also expanded as 
\begin{eqnarray}
\Delta_{\zeta_1\zeta_2}(k) &=&
\Delta_{\zeta_1\zeta_2}^{(1)}(k)+
\Delta_{\zeta_1\zeta_2}^{(2)}(k)+
\Delta_{\zeta_1\zeta_2}^{(3-1)}(k) \nonumber\\
&& + \Delta_{\zeta_1\zeta_2}^{(3-2)}(k)+
\Delta_{\zeta_1\zeta_2}^{(3-2*)}(k)+
\Delta_{\zeta_1\zeta_2}^{(3-3)}(k)+
\Delta_{\zeta_1\zeta_2}^{(3-3*)}(k). 
\label{eq:Delta}
\end{eqnarray}
The analytic expressions of the contributions are 
\begin{eqnarray}
\Delta_{\zeta_1\zeta_2}^{(1)}(k) &=& 
- \frac{1}{2}\frac{T}{N}
\sum_{k'}\sum_{\zeta_3\zeta_4}
F_{\zeta_3\zeta_4}(k')
\Gamma_{\zeta_1\zeta_2, \zeta_4\zeta_3}^{(0)},
\label{eq:Delta(1)}\\
\Delta_{\zeta_1\zeta_2}^{(2)}(k) &=&
\frac{T}{N} \sum_{k'}\sum_{\zeta_3\zeta_4}
\sum_{\{\gamma_n\}}
F_{\zeta_3\zeta_4}(k')
\chi_{\gamma_4\gamma_2,\gamma_1\gamma_3}^{(0)}(k-k') 
\Gamma_{\zeta_1\gamma_4, \gamma_2\zeta_3}^{(0)} 
\Gamma_{\gamma_1\zeta_2, \zeta_4\gamma_3}^{(0)},
\label{eq:Delta(2)}\\
\Delta_{\zeta_1\zeta_2}^{(3-1)}(k) &=& 
- \frac{T}{N} \sum_{k'}\sum_{\zeta_3\zeta_4}
\sum_{\{\gamma_n\}} \sum_{\{\xi_n\}}
F_{\zeta_3\zeta_4}(k')
\chi_{\gamma_4\gamma_2,\gamma_1\gamma_3}^{(0)}(k-k') 
\chi_{\xi_4\xi_2,\xi_1\xi_3}^{(0)}(k-k') \nonumber\\
&& \times \Gamma_{\zeta_1\gamma_4, \gamma_2\zeta_3}^{(0)}
\Gamma_{\gamma_1\xi_4, \xi_2\gamma_3}^{(0)}
\Gamma_{\xi_1\zeta_2, \zeta_4\xi_3}^{(0)},
\label{eq:Delta(3-1)}\\
\Delta_{\zeta_1\zeta_2}^{(3-2)}(k) &=&
- \biggl( \frac{T}{N} \biggr)^2
\sum_{k'k_1}\sum_{\zeta_3\zeta_4}
\sum_{\{\gamma_n\}} \sum_{\{\xi_n\}}
F_{\zeta_3\zeta_4}(k') G_{\gamma_2\gamma_1}^{(0)}(k-k'+k_1)
\chi_{\xi_4\xi_1,\xi_2\xi_3}^{(0)}(k+k_1) \nonumber\\
&& \times G_{\gamma_3\gamma_4}^{(0)}(k_1)
\Gamma_{\zeta_1\gamma_4, \gamma_2\zeta_3}^{(0)}
\Gamma_{\gamma_1\xi_3, \zeta_4\xi_2}^{(0)}
\Gamma_{\xi_1\zeta_2, \xi_4\gamma_3}^{(0)},
\label{eq:Delta(3-2)}\\
\Delta_{\zeta_1\zeta_2}^{(3-2*)}(k) &=&
- \biggl( \frac{T}{N} \biggr)^2
\sum_{k'k_1}\sum_{\zeta_3\zeta_4} 
\sum_{\{\gamma_n\}} \sum_{\{\xi_n\}}
F_{\zeta_3\zeta_4}(k')
G_{\gamma_1\gamma_2}^{(0)}(-k+k'+k_1) 
\chi_{\xi_3\xi_2,\xi_1\xi_4}^{(0)}(-k+k_1) \nonumber\\
&& \times G_{\gamma_4\gamma_3}^{(0)}(k_1)
\Gamma_{\zeta_1\xi_2, \gamma_4\xi_3}^{(0)}
\Gamma_{\xi_4\gamma_2, \xi_1\zeta_3}^{(0)}
\Gamma_{\gamma_3\zeta_2, \zeta_4\gamma_1}^{(0)},
\label{eq:Delta(3-2*)}\\
\Delta_{\zeta_1\zeta_2}^{(3-3)}(k) &=&
- \frac{1}{2} \biggl( \frac{T}{N} \biggr)^2 
\sum_{k'k_1}\sum_{\zeta_3\zeta_4}
\sum_{\{\gamma_n\}} \sum_{\{\xi_n\}}
F_{\zeta_3\zeta_4}(k') 
G_{\gamma_1\gamma_2}^{(0)}(-k+k'+k_1) 
\phi_{\xi_1\xi_4,\xi_2\xi_3}^{(0)}(-k+k_1) \nonumber\\
&& \times G_{\gamma_4\gamma_3}^{(0)}(k_1)
\Gamma_{\zeta_1\gamma_2, \gamma_4\zeta_3}^{(0)}
\Gamma_{\xi_3\xi_2, \zeta_4\gamma_1}^{(0)}
\Gamma_{\gamma_3\zeta_2, \xi_1\xi_4}^{(0)},
\label{eq:Delta(3-3)}\\
\Delta_{\zeta_1\zeta_2}^{(3-3*)}(k) &=&
- \frac{1}{2} \biggl( \frac{T}{N} \biggr)^2 
\sum_{k'k_1}\sum_{\zeta_3\zeta_4}
\sum_{\{\gamma_n\}} \sum_{\{\xi_n\}}
F_{\zeta_3\zeta_4}(k')
G_{\gamma_2\gamma_1}^{(0)}(k-k'+k_1)
\phi_{\xi_2\xi_3,\xi_1\xi_4}^{(0)}(k+k_1) \nonumber\\
&& \times G_{\gamma_3\gamma_4}^{(0)}(k_1)
\Gamma_{\zeta_1\gamma_4, \xi_3\xi_2}^{(0)}
\Gamma_{\xi_1\xi_4, \gamma_2\zeta_3}^{(0)}
\Gamma_{\gamma_1\zeta_2, \zeta_4\gamma_3}^{(0)}, 
\label{eq:Delta(3-3*)}
\end{eqnarray}
with 
\begin{eqnarray}
\chi_{\xi_1\xi_2,\xi_3\xi_4}^{(0)}(q)= 
- \frac{T}{N}\sum_k G_{\xi_4\xi_1}^{(0)}(k) G_{\xi_2\xi_3}^{(0)}(q+k),\\ 
\phi_{\xi_1\xi_2,\xi_3\xi_4}^{(0)}(q)= 
- \frac{T}{N}\sum_k G_{\xi_1\xi_3}^{(0)}(k) G_{\xi_2\xi_4}^{(0)}(q-k). 
\end{eqnarray} 
The factor of $\frac{1}{2}$ in eqs.~(\ref{eq:Delta(1)}), 
(\ref{eq:Delta(3-3)}), and (\ref{eq:Delta(3-3*)}) 
is necessary to avoid double counting. 
The band indices are replaced with orbital indices 
by using the diagonalization matrix: 
\begin{eqnarray}
\Delta_{a\sigma_1\sigma_2}(k) = 
\sum_{\ell_1\ell_2} 
U_{\ell_1 a}^*({\mib k})U_{\ell_2 a}^*(-{\mib k})
\Delta_{\zeta_1\zeta_2}(k), 
\label{eq:DeltaDelta}\\
F_{\zeta_1\zeta_2}(k) = 
\sum_a U_{\ell_1 a}({\mib k})U_{\ell_2 a}(-{\mib k})
F_{a\sigma_1\sigma_2}(k). 
\label{eq:FF}
\end{eqnarray}
The anomalous Green's function is related to the anomalous 
self-energy by 
\begin{equation}
F_{a\sigma_1\sigma_2}(k) = |G_a^{(0)}(k)|^2 \Delta_{a\sigma_1\sigma_2}(k)
\label{eq:Gorkov} 
\end{equation}
in the band representation. 
Note that this relation is valid only in the region 
where the superconducting order parameter is nonzero but small 
(i.e., in the vicinity of transition points), 
since all quantities in the present formulation 
are linearized with respect to the anomalous functions. 
It is clear that eq.~(\ref{eq:Eliashberg}) 
can be consistently obtained from eqs.~(\ref{eq:Eliashberg2}), 
(\ref{eq:FF}), (\ref{eq:Gorkov}), (\ref{eq:DeltaDelta}), 
and (\ref{eq:Vaa}). 
The anomalous functions $\Delta_{a\sigma_1\sigma_2}(k)$ 
and $F_{a\sigma_1\sigma_2}(k)$ can be determined 
by eqs.~(\ref{eq:Eliashberg}) and (\ref{eq:Gorkov}), 
which is carried out numerically. 
For numerical calculations in the present study, 
we take 32 $\times$ 32 ${\mib k}$-points 
in the first Brillouin zone and 512 Matsubara frequencies. 
The momentum and frequency summations are performed 
using the fast Fourier transformation. 

We now turn our attention to the remaining degrees of freedom 
in the phase factor of the diagonalization matrix elements 
$U_{\ell a}({\mib k})$. We can easily verify that 
if $U_{\ell a}({\mib k})$ diagonalizes $H_0$, 
then $U_{\ell a}({\mib k})e^{{\rm i} \phi_a({\mib k})}$ 
also diagonalizes $H_0$, where $\phi_a({\mib k})$ 
is an arbitrary real function of ${\mib k}$. 
It is easy to see that $G_{\zeta_1\zeta_2}^{(0)}(k)$ 
and $V_{\zeta_1\zeta_2,\zeta_3\zeta_4}(k, k')$ do not depend 
on the choice of phase function $\phi_a({\mib k})$. 
However, the values of $V_{a\sigma_1\sigma_2,a'\sigma_3\sigma_4}(k, k')$ 
and the anomalous functions depend on $\phi_a({\mib k})$ in general, 
since they are subject to extrinsic phase modulations. 
In order to avoid this unfavorable situation, 
it is natural to assume the following condition: 
\begin{equation}
\label{eq:UU}
U_{\ell a}({\mib k}) = U_{\ell a}^*(-{\mib k}). 
\end{equation}
We use this condition throughout the present work. 

In the present formulation, we have taken only the diagonal components 
of anomalous functions with respect to band indices 
and have neglected the off-diagonal components. 
Here note that the diagonal components describe intraband pairing, 
while the off-diagonal ones describe interband pairing. 
Superconducting instability is generally connected 
with the logarithmic increase in the pair-correlation function 
at low temperatures. While the diagonal pair correlations 
exhibit this increase down to zero temperature, 
the off-diagonal ones do not in general. 
This is because the increase of the pair correlations 
between different bands stops below a temperature 
comparable to the energy difference between the bands. 
Thus, we may neglect the off-diagonal components for the reliable 
prediction of pairing symmetry at low temperatures. 

The same formulation presented here was previously applied 
to the spin-triplet superconductivity in Sr$_2$RuO$_4$ 
by the present author and Yamada~\cite{ref:NomuraT2002}. 
In the present section we have given improved 
and more useful analytic expressions 
for the perturbation terms. 

\section{Analysis of 5-band Hubbard model}
\label{sc:5-band analysis}

\subsection{5-band tight-binding model by Kuroki et al.}
\label{sc:5-band}

In this section, we present an analysis based on a 5-band model.   
Following Kuroki et al~\cite{ref:KurokiK2008}, 
we take the five local Fe3d-like orbitals: 
$\ell = 3Z^2-R^2$, $XZ$, $YZ$, $X^2-Y^2$, and $XY$. 
The tight-binding parameters $t_{i\ell, j\ell'}$ 
are provided in ref.~\ref{ref:KurokiK2008}. 
Here we should bear in mind that the original unit cell 
contains two iron atoms. 
The principal axes $X$ and $Y$ are taken 
along the basal plane parallel to the FeAs layers, and 
the $Z$-axis is taken to be perpendicular to the basal plane, 
with respect to this original unit cell. 
The above orbital states are represented 
using these original coordinates. 
Neglecting the As sites in the effective models, 
the remaining iron atoms form a two-dimensional square lattice 
in which each unit cell contains only one iron atom. 
The principal axes $x$ and $y$ with respect to this new unit cell 
are parallel to the nearest-neighbor iron-iron bonds 
and are obtained by rotating the original axes $X$ and $Y$ 
by 45$^\circ$ in the basal plane. 
In momentum space, the original Brillouin zone in which 
positions are parametrized by ${\mib k} = (k_X, k_Y)$ is extended 
to a wide Brillouin zone in which positions are parametrized 
by ${\mib k} = (k_x, k_y)$. 

\begin{figure}
\begin{center}
\includegraphics[width=0.9\linewidth]{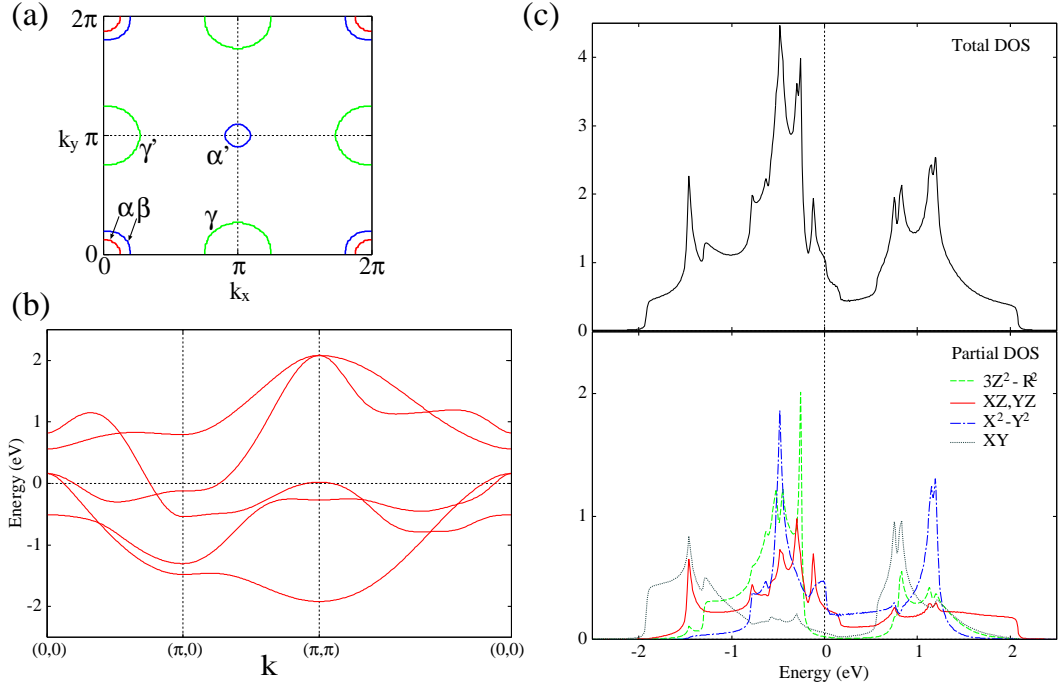}
\end{center}
\caption{
(Color online) 
Electronic structure of the 5-band model by Kuroki et al. 
in ref.~\ref{ref:KurokiK2008}. 
(a) Fermi surface and (b) energy band dispersion, 
both of which are depicted in the extended Brillouin zone. 
(c) total density of states and partial density of states 
for the orbitals, $3Z^2-R^2$, $XZ$, $YZ$, $X^2-Y^2$ and $XY$. 
The Fermi level is set to zero on the horizontal axis. 
}
\label{model5b}
\end{figure}

The electronic structure of the 5-band model is shown in Fig.~\ref{model5b}. 
The total electron number $n$ is related to the doping level $x$ by $x=n-6$. 
The Fermi surface for $x=0.1$ forms five Fermi pockets 
in the extended Brillouin zone: 
two hole pockets ($\alpha$ and $\beta$) around $(0, 0)$, 
one hole pocket ($\alpha'$) around $(\pi, \pi)$, 
and electron pockets ($\gamma$ and $\gamma'$) 
around $(\pi, 0)$ and $(0,\pi)$. 
In the original folded representation, 
both $(0, 0)$ and $(\pi, \pi)$ in the extended Brillouin zone 
are folded onto the $\Gamma$ point, 
and $(\pi, 0)$ and $(0, \pi)$ are folded onto the M point. 
Thus, three concentric hole-like surfaces will be located 
around the $\Gamma$ point, and two concentric electron-like surfaces 
will be constructed around the M point, 
in the original folded Brillouin zone. 
The $XZ$, $YZ$, and $X^2-Y^2$ states occupy most of the density 
of states near the Fermi level and are naturally expected 
to dominate the electronic properties including the superconductivity. 

\subsection{Favorable pairing symmetry}
\label{sc:Favorable pairing}

\begin{figure} 
\begin{center}
\includegraphics[width=0.8\linewidth]{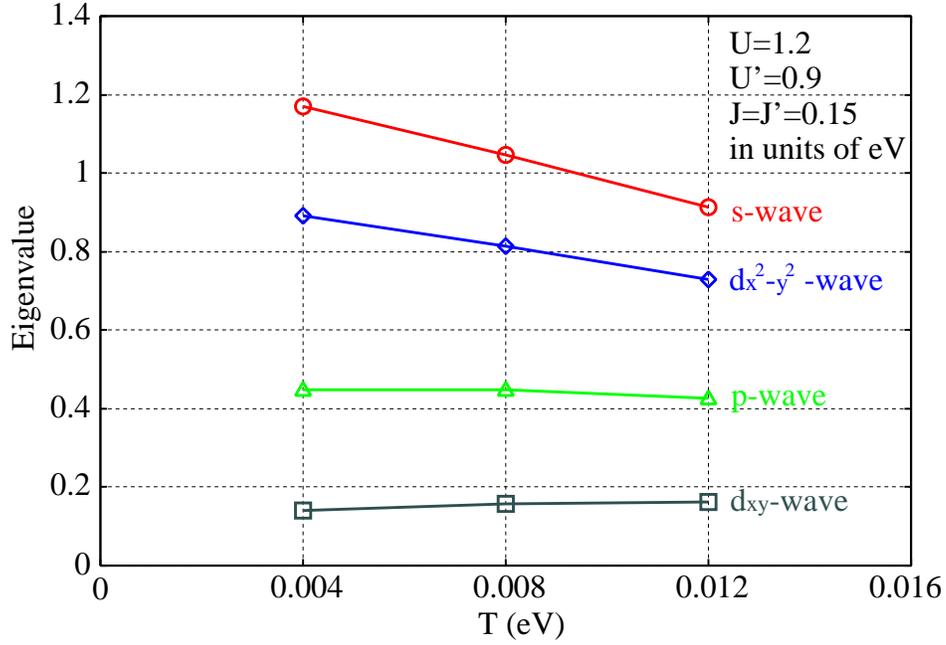}
\end{center}
\caption{
(Color online)
Eigenvalues of the Eliashberg equation as a function 
of temperature for various pairing symmetries\cite{cmnt:NomuraT}. 
The doping level is $x=0.10$. 
}
\label{egnvs5b_2}
\end{figure}

The most favorable pairing symmetry is obtained from the eigenfunction 
giving the maximum eigenvalues of the Eliashberg equation. 
In Fig.~\ref{egnvs5b_2}, the eigenvalues 
for various pairing symmetries at $x=0.10$ are shown 
as a function of temperature. 
The most favorable pairing symmetry is the spin-singlet $s$-wave 
(or more appropriately, as we shall see below, 
the extended $s$-wave or $s_\pm$-wave), 
and the maximum eigenvalue is sufficiently large 
to explain the actual high transition temperatures 
for realistic values of the Coulomb interaction: 
$U=1.2$ eV, $U'=0.9$ eV, and $J=J'=0.15$ eV. 
In Fig.~\ref{egnvs5b_2}, the transition temperature is evaluated 
to be about 100 K from the condition $\lambda_{\rm max}(T_{\rm c})=1$. 

\begin{figure}
\begin{center}
\includegraphics[width=\linewidth]{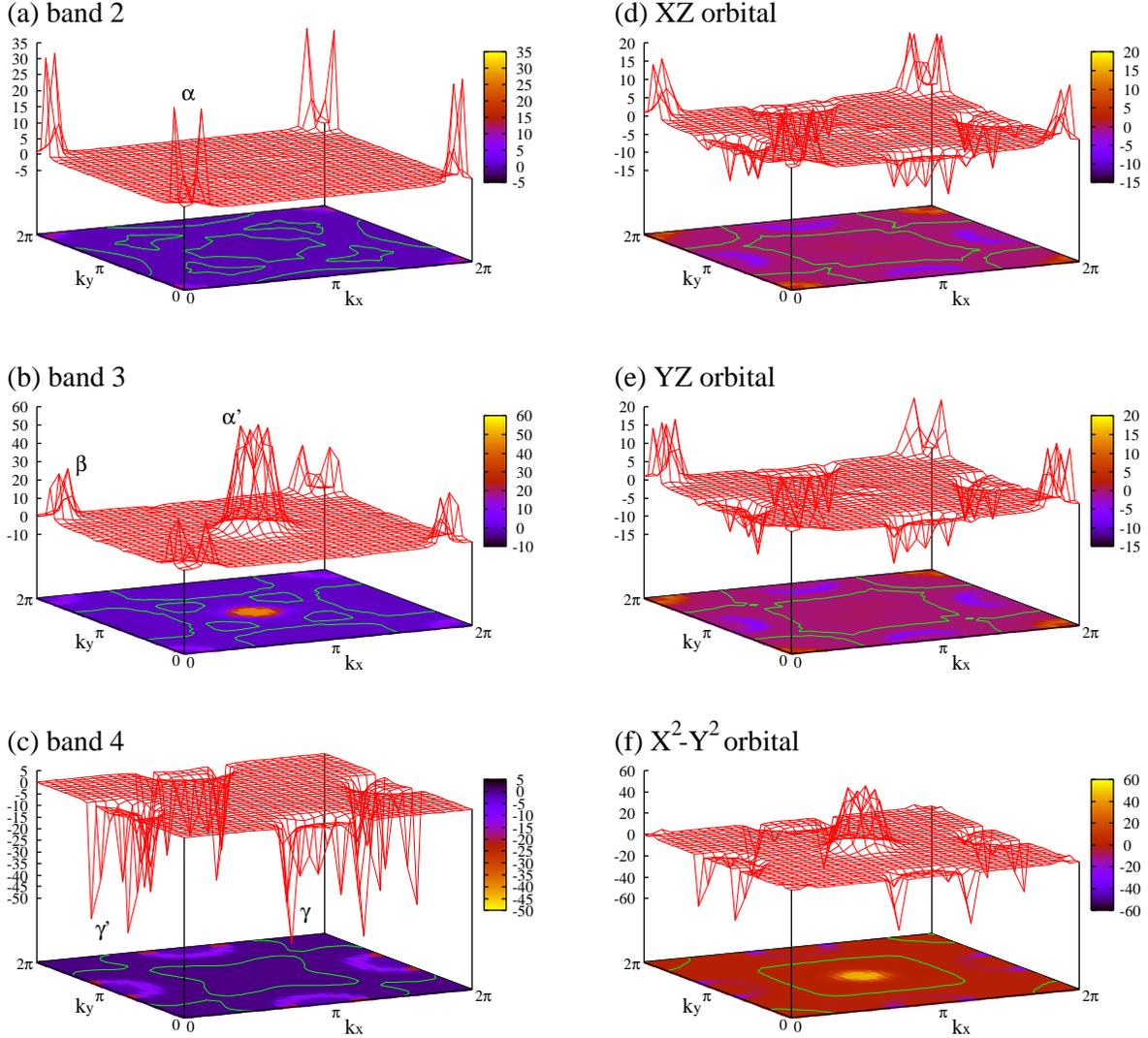}
\end{center}
\caption{
(Color online)
Momentum dependence of the anomalous Green's functions 
at $x=0.10$ and $T=$0.008 eV in the extended Brillouin zone. 
(a) $F_2(\mib{k}, {\rm i} \pi T)$, (b) $F_3(\mib{k}, {\rm i} \pi T)$, 
(c) $F_4(\mib{k}, {\rm i} \pi T)$, (d) $F_{XZ\,XZ}(\mib{k}, {\rm i} \pi T)$, 
(e) $F_{YZ\,YZ}(\mib{k}, {\rm i} \pi T)$, and 
(f) $F_{X^2-Y^2\,X^2-Y^2}(\mib{k}, {\rm i} \pi T)$. 
Note that (a)-(c) show $F_a(k)$ ($a=2,3,4$) 
in the diagonalized-band representation, 
while (d)-(f) show three diagonal elements $F_{\ell\ell}({\mib k})$ 
($\ell = XZ$, $YZ$, $X^2-Y^2$) in the orbital representation. 
The indices $\alpha$, $\alpha$', $\beta$, $\gamma$, and $\gamma$' 
in (a)-(c) denote the Fermi pockets shown in Fig.~\ref{model5b}(a). 
The green solid lines on the base plane indicate 
the zero contours of the anomalous Green's functions. 
We find in (a)-(c) that the green lines do not cross 
the Fermi surfaces, i.e., there are no nodes 
of the superconducting order parameter on the Fermi surfaces.
}
\label{Fktot}
\end{figure}

We number the energy bands $E_a({\mib k})$ ($a=1,\cdots, 5$) 
in ascending order with respect to energy, i.e., 
$a<a'$ for $E_a({\mib k}) < E_{a'}({\mib k})$ 
at any ${\mib k}$. 
The three bands, $a=2,3$, and 4, cross the Fermi level. 
The momentum dependences of the anomalous Green's functions $F_a(k)$ 
at $\omega_n = \pi T$ on these three bands 
are displayed in Figs.~\ref{Fktot}(a)-(c). 
Here we have suppressed the spin part of the spin-singlet anomalous 
Green's functions by using the Pauli spin matrix $\sigma_y$: 
$F_{a\sigma_1\sigma_2}(k)=F_a(k)[{\rm i}\sigma_y]_{\sigma_1\sigma_2}$. 
As we see in Fig.~\ref{Fktot}, the zero contours of the anomalous functions 
do not have any cross sections with the Fermi surfaces. 
Thus, we conclude that the superconducting gap will fully open 
on the Fermi surfaces, noting that the nodal positions 
of the superconducting gap generally coincide with those 
of the anomalous self-energy. 
A remarkable feature is the sign change of the superconducting order 
parameter between the hole pockets ($\alpha$, $\alpha'$ and $\beta$) 
and the electron pockets ($\gamma$ and $\gamma'$). 
Thus, the pairing symmetry is not a simple $s$-wave, 
but an extended $s$-wave symmetry, 
or more specifically, an $s_\pm$-wave symmetry. 

The superconducting order parameter on the inner hole pockets 
$\alpha$ and $\alpha'$ around $(0, 0)$ takes nearly 
the same absolute value as that on the electron pockets 
$\gamma$ and $\gamma'$ around $(\pi, 0)$ and $(0, \pi)$, 
although the sign of the order parameter is changed between them. 
The absolute magnitude of the order parameter on the outer hole 
pocket $\beta$ is about half of that on the inner hole pockets 
$\alpha$ and $\alpha'$. These features are in agreement 
with the results of ARPES experiments~\cite{ref:DingH2008}. 
Numerical calculations with finer ${\mib k}$-meshes will be necessary 
to investigate the detailed momentum dependence of the order parameter 
on each small Fermi circle. 

We turn our attention to the anomalous Green's functions 
in the orbital representation, $F_{\ell_1\ell_2}(k)$, 
which are obtained from $F_a(k)$ by eq.~(\ref{eq:FF}). 
We display only three diagonal components 
in the orbital representation, $F_{XZ\,XZ}(k)$, 
$F_{YZ\,YZ}(k)$, and $F_{X^2-Y^2\, X^2-Y^2}(k)$, 
at $\omega_n = \pi T$, 
since the other remaining diagonal elements are much smaller. 
Note that not all of the off-diagonal components 
in the orbital representation are so small, for example, $F_{XZ\,YZ}(k)$ 
and $F_{YZ\,XZ}(k)$ (whose plots are not shown in the present article) 
have magnitudes comparable to $F_{XZ\,XZ}(k)$ and $F_{YZ\,YZ}(k)$. 
This is due to the fact that these orbitals are well hybridized 
with each other. $F_{XZ\,X^2-Y^2}(k)$, $F_{YZ\,X^2-Y^2}(k)$, 
$F_{X^2-Y^2\,XZ}(k)$, and $F_{X^2-Y^2\,YZ}(k)$ 
(whose plots are not shown in the present article) are negligibly small. 
An important feature of the anomalous Green's function 
in the orbital representation is that, roughly speaking, 
the maximum and minimum of $F_{\ell\ell}(k)$ in momentum space 
have nearly the same absolute magnitude but opposite signs 
to each other, and consequently $\sum_{\mib k} F_{\ell\ell}(k)$, 
which indicates the strength of the on-site pair correlation at orbital $\ell$, 
is suppressed to a small value for each $\ell$. 
We stress that this is not an accidental feature 
but is essential for the unconventional $s_\pm$-wave pairing 
due to electron correlations, as we shall discuss later 
in \S\ref{sc:Discussion}. 

\subsection{Doping dependence of transition temperature}
\label{sc:Doping dependence}

\begin{figure}
\begin{center}
\includegraphics[width=0.8\linewidth]{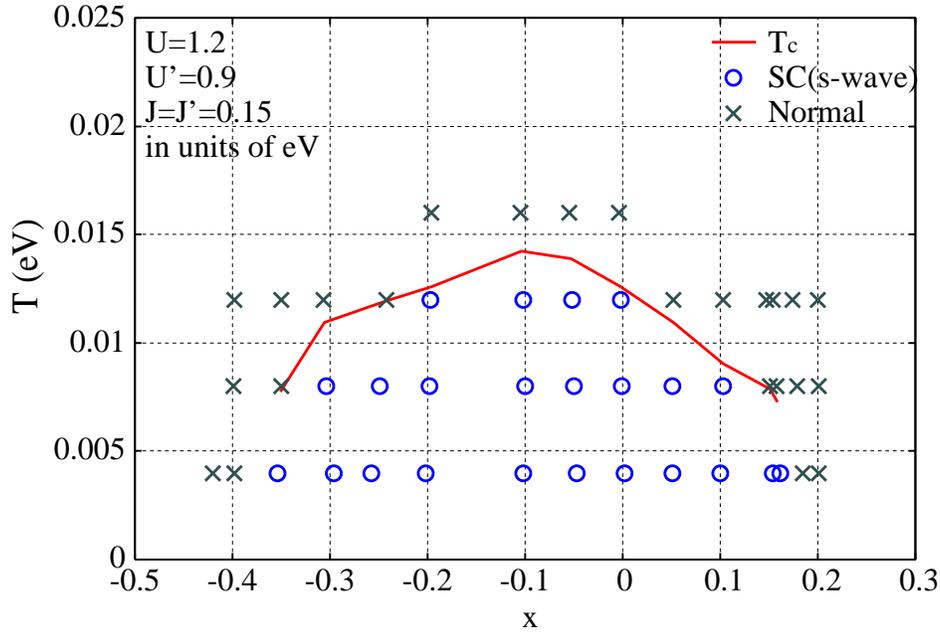}
\end{center}
\caption{
(Color online)
Phase diagram. The horizontal and vertical axes 
denote electron doping level $x$ and temperature (in units of eV). 
Negative values of $x$ indicate hole doping. 
Circles (crosses) denote that the eigenvalue 
of the Eliashberg equation is larger (smaller) than unity, 
i.e, $\lambda_{\rm max}>1$ ($\lambda_{\rm max}<1$). 
In other words, the circles and crosses 
represent superconducting and normal states, respectively. 
The solid curve represents the transition temperature 
determined by linear interpolation. 
The Coulomb integrals are $U=1.2$ eV, $U'=0.9$ eV, 
and $J=J'=0.15$ eV. 
}
\label{tcx3}
\end{figure}

Experiments suggest that carrier doping is an important method 
for controlling iron-pnictide superconductivity. 
We investigate the effects of carrier doping 
on the transition temperature and pairing symmetry. 
In the present study, we assume simply that carrier doping, 
which is realized by the chemical substitution 
of atoms in experiments, only shifts the chemical potential 
and retains rigidly the band structure. 
The dependence of the transition temperature on doping level $x$ 
is displayed in Fig.~\ref{tcx3}, where the transition temperature 
is determined by the linear interpolation of eigenvalues. 
We find that the superconducting state also extends 
in the hole-doped region $x<0$. 
Thus, our theory is consistent with the existence 
of superconductivity in hole-doped cases, 
which was suggested by some experiments. 
Although in practice the structural transition and SDW ordering 
inhibit the superconductivity near the undoped case $x=0$, 
the structural transitions and the SDW ordering 
are unfortunately beyond the scope of our theory. 

\begin{figure}
\begin{center}
\includegraphics[width=0.8\linewidth]{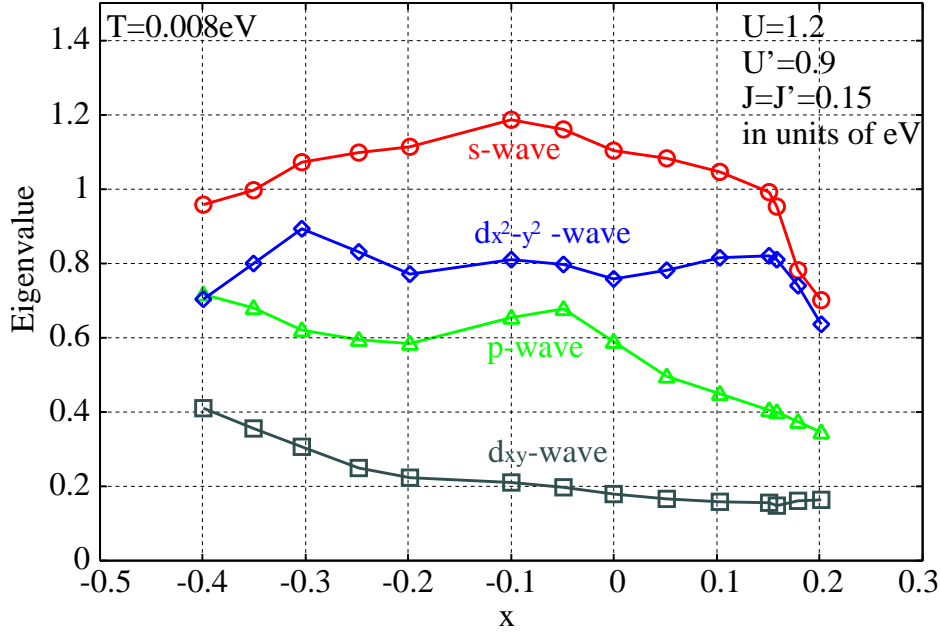}
\end{center}
\caption{
(Color online)
Doping dependence of the eigenvalues of the Eliashberg equation 
for $U=1.2$ eV, $U'=0.9$ eV and $J=J'=0.15$ eV at $T=0.008$ eV. 
}
\label{egnvsx}
\end{figure}

Concerning the pairing symmetry, 
the spin-singlet $s_\pm$-wave symmetry does not change 
over the whole doping region that we have calculated. 
The eigenvalues at $T=0.008$ K for various symmetries 
are plotted in Fig.~\ref{egnvsx}. 
The $s_\pm$-wave state is likely to occur robustly 
against other pairing symmetries in iron pnictides. 
Varying the doping level, the volumes of the Fermi pockets 
are changed, but the zero contours of the anomalous functions 
always avoid crossing the Fermi surfaces. Thus, the fully gapped 
state is realized over the whole superconducting region. 
Here we note that the transition temperature is suddenly suppressed 
at a high doping level on both sides of the electron- and hole-doping regions. 
This is because the hole pocket $\alpha'$ around $(\pi,\pi)$ disappears 
near this electron-doping level and the electron pockets $\gamma$ 
and $\gamma'$ around $(\pi,0)$ and $(0,\pi)$ disappear 
near this hole-doping level. 

\subsection{Effect of vertex corrections}
\label{sc:Vertex corrections}

\begin{figure}
\begin{center}
\includegraphics[width=0.8\linewidth]{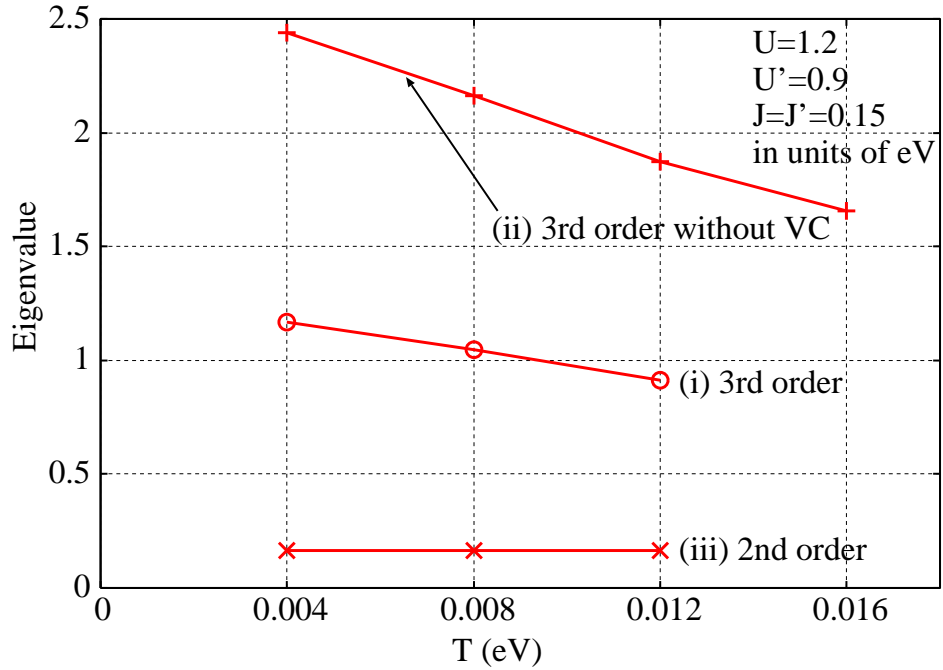}
\end{center}
\caption{
(Color online)
Eigenvalues of the Eliashberg equation for three cases: 
(i) third-order calculation (including the vertex corrections), 
(ii) third-order calculation without the vertex corrections, 
and (iii) second-order calculation. 
The Coulomb integrals are $U=1.2$ eV, $U'=0.9$ eV 
and $J=J'=0.15$ eV. 
}
\label{egnvs5b_vc}
\end{figure}

It will be interesting to study the effects of the vertex corrections 
on the eigenvalues of the Eliashberg equation. 
The eigenvalues are compared for the following three cases: 
(i) the third-order calculation including the vertex corrections, 
using eq.~(\ref{eq:Delta}), 
(ii) the third-order calculation without the vertex corrections, 
i.e., using $\Delta_{\zeta_1\zeta_2}(k)=\Delta_{\zeta_1\zeta_2}^{(1)}(k)+
\Delta_{\zeta_1\zeta_2}^{(2)}(k)+\Delta_{\zeta_1\zeta_2}^{(3-1)}(k)$
instead of eq.~(\ref{eq:Delta}), and 
(iii) the second-order calculation, i.e., using 
$\Delta_{\zeta_1\zeta_2}(k)=\Delta_{\zeta_1\zeta_2}^{(1)}(k)+
\Delta_{\zeta_1\zeta_2}^{(2)}(k)$ instead of eq.~(\ref{eq:Delta}).
The calculated results are displayed in Fig.~\ref{egnvs5b_vc}. 
We see that the second-order theory cannot quantitatively 
explain the high $T_{\rm c}$ within the realistic strength 
of the Coulomb interaction, because the eigenvalues are too small. 
The vertex corrections reduce the eigenvalues, 
and suppress the transition temperature. 
Thus, we consider that transition temperatures 
are overestimated quantitatively, 
when using only the terms included in the RPA. 
This is qualitatively the same as in the $d$-wave superconductivity 
of single-band systems such as cuprates 
and organics~\cite{ref:HottaT1994,ref:JujoT1999,ref:NomuraT2001}. 
The momentum dependence of the anomalous Green's functions $F_a(k)$ 
calculated without the vertex corrections 
(whose plots are not shown in the present article) 
is almost the same as that calculated with the vertex corrections. 
We consider that the vertex corrections do not 
affect the momentum dependence of $F_a(k)$ qualitatively, 
although they lower the transition temperature. 

\subsection{Dependence on Coulomb integrals}
\label{sc:Coulomb dependence}

Eigenvalues of the Eliashberg equation for various sets 
of Coulomb integrals are displayed in Table~\ref{tabegnvs}. 
We find that the $s$-wave ($s_\pm$-wave) state is the most favorable 
pairing state robustly against the other pairing states. 
The order of pairing strength is $s_\pm > d_{x^2-y^2} > p > d_{xy}$ 
for all the cases of Coulomb integrals that we have calculated. 
In Table~\ref{tbl:Eigenvalues}, comparing the upper 
three cases of $U=1.2$, we speculate 
that eigenvalues become small for small $U'$. 
We infer from this that processes including 
the interorbital Coulomb scattering $U'$ 
play a role of enhancing the eigenvalues. 
By comparison with the last line 
in Table~\ref{tbl:Eigenvalues}, 
we find that the eigenvalues are reduced 
by using a small $U'$, but are recovered 
by taking slightly larger $U$. 
\begin{table}[t]
\caption{
Eigenvalues of the Eliashberg equation 
for various Coulomb integrals. $x=0.10$ and $T=0.004$ eV.}
\label{tabegnvs}
\begin{tabular}{|l|c c c c|}
\hline
Coulomb integrals (eV) & $s$-wave & $p$-wave 
& $d_{x^2-y^2}$-wave & $d_{xy}$-wave \\ \hline 
$U=1.2$, $U'=0.9$, $J=J'=0.15$ & 1.17 & 0.45 & 0.89 & 0.14 \\
$U=1.2$, $U'=0.6$, $J=J'=0.3$ & 0.56 & 0.068 & 0.39 & 0.052 \\
$U=1.2$, $U'=0.9$, $J=J'=0$ & 1.34 & 0.58 & 1.05 & 0.18 \\
$U=1.5$, $U'=0.5$, $J=J'=0.5$ & 1.09 & 0.19 & 0.71 & 0.12 \\
\hline
\end{tabular}
\label{tbl:Eigenvalues}
\end{table}

\section{Analysis of 2-band Hubbard model}
\label{sc:2-band analysis}
\subsection{2-band tight-binding model by Raghu et al.}
\label{sc:2-band}

\begin{figure}
\begin{center}
\includegraphics[width=0.9\linewidth]{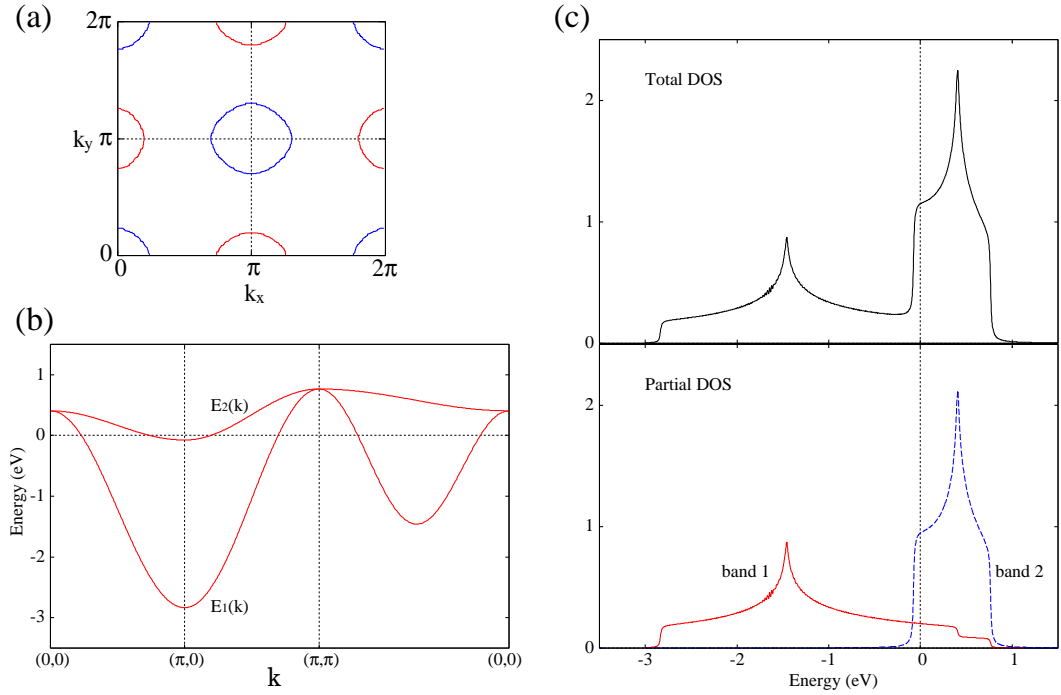}
\end{center}
\caption{
(Color online)
Electronic structure of the 2-band model 
by Raghu et al. in ref.~\ref{ref:RaghuS2008}.
(a) Fermi surface, (b) energy band dispersion, 
and (c) density of states. 
}
\label{model2b}
\end{figure}

We now use another simpler model. 
Raghu et al. proposed a two-band model, taking account of 
only the $xz$ and $yz$ orbitals~\cite{ref:RaghuS2008}: 
\begin{eqnarray}
E_{xz}({\mib k}) &=& 2 t_1 \cos k_x + 2 t_2 \cos k_y + 4 t_3 \cos k_x \cos k_y, \\
E_{yz}({\mib k}) &=& 2 t_2 \cos k_x + 2 t_1 \cos k_y + 4 t_3 \cos k_x \cos k_y, \\
E_{\rm hyb}({\mib k}) &=& - 4 t_4 \sin k_x \sin k_y, 
\end{eqnarray}
where the momentum variables $k_x$ and $k_y$ represent 
positions in the extended Brillouin zone, 
and the orbitals $xz$ and $yz$ should not be confused 
with the $XZ$ and $YZ$ orbitals in the original representation. 
$E_{\rm hyb}({\mib k})$ is the hybridization term 
between the $xz$ and $yz$ bands. 
We take $t_1=0.3$, $t_2=-0.39$, $t_3=0.255$, and $t_4=-0.255$, 
in units of eV, which give a quantitatively realistic energy scale, 
specifically, a total bandwidth of about 4 eV. 
The chemical potential is $\mu=0.435$ eV. 
The diagonalized band energies are 
\begin{eqnarray}
E_1({\mib k}) = E_+({\mib k}) 
- \sqrt{E_-({\mib k})^2 + E_{\rm hyb}({\mib k})^2}, \\
E_2({\mib k}) = E_+({\mib k}) 
+ \sqrt{E_-({\mib k})^2 + E_{\rm hyb}({\mib k})^2}, 
\end{eqnarray}
with 
\begin{equation}
E_\pm({\mib k}) = \frac{E_{xz}({\mib k}) \pm E_{yz}({\mib k})}{2}. 
\end{equation}
The electronic structure of this model is given in Fig.~\ref{model2b}. 
The Fermi surface is qualitatively obtained from that 
by electronic band structure calculations 
and ARPES experiments, i.e., electron pockets 
around $(0, 0)$ and $(\pi, \pi)$ 
and hole pockets around $(\pi, 0)$ and $(0, \pi)$. 
However, the energy band dispersions and the density of states 
are markedly different from those obtained from band calculations. 

\subsection{Eigenvalues of Eliashberg equation}
\label{sc:2-band eingenvalues}

\begin{figure}
\begin{center}
\includegraphics[width=0.8\linewidth]{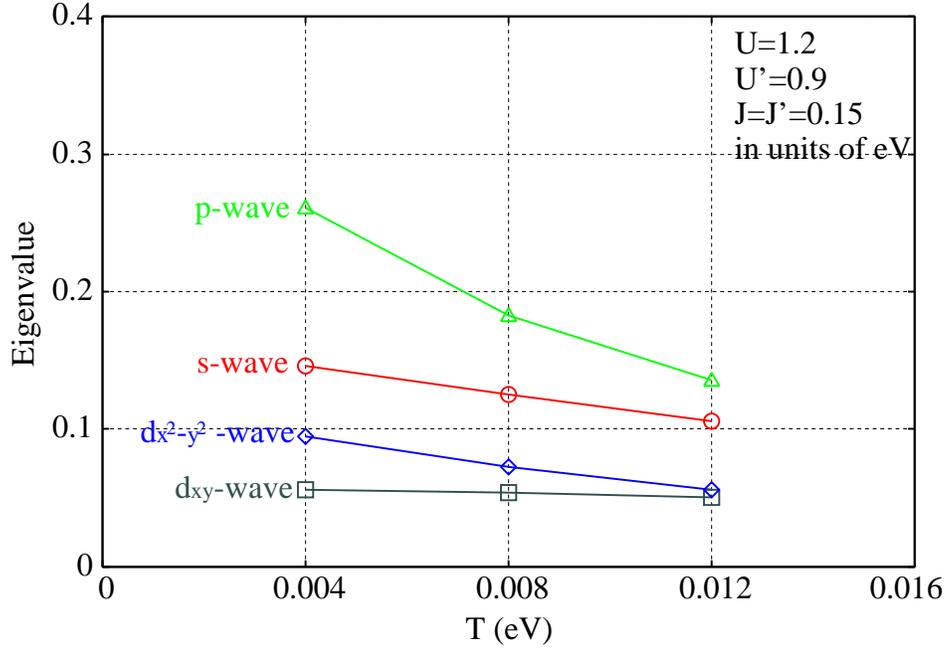}
\end{center}
\caption{
(Color online)
Eigenvalues of Eliashberg equation 
for the 2-band model as a function of temperature. 
The Coulomb integrals are $U=1.2$ eV, $U'=0.9$ eV, 
and $J=J'=0.15$ eV.} 
\label{egnvs2b}
\end{figure}

For this simple 2-band model, we perform the same analysis as 
that for the 5-band model in \S\ref{sc:5-band analysis}. 
The calculated eigenvalues of the Eliashberg equation 
are presented in Fig.~\ref{egnvs2b}. 
The most favorable pairing symmetry is the odd-parity spin-triplet $p$-wave symmetry. 
This is due to the effect of the vertex corrections. 
As discussed in the case of Sr$_2$RuO$_4$~\cite{ref:NomuraT2000}, 
the influence of the vertex corrections will be strong 
for such large particle-hole asymmetry as that 
in the present electronic structure. 
However the obtained eigenvalues are too small to explain 
the high $T_{\rm c}$, and the triplet pairing state 
is not expected to be realized. 
Thus, we infer that the 2-band model is insufficient 
to describe the high-$T_{\rm c}$ superconductivity 
in iron pnictides. 

\section{Discussion}
\label{sc:Discussion}

We have seen in \S\ref{sc:Favorable pairing} 
that the eigenvalues of the Eliashberg equation 
for the 5-band model are sufficiently large to explain 
the high transition temperatures. 
The highest transition temperature that can be expected 
according to Fig.~\ref{tcx3} is about 160 K at around $x=-0.1$, 
and higher than the observed highest value of 56 K. 
It is, of course, possible to obtain more realistic values 
of $T_{\rm c}$ by taking slightly smaller values 
of the Coulomb integrals. 
In the present calculations, we have not included 
the normal self-energy corrections in the Green's functions. 
If we include these corrections, then $T_{\rm c}$ 
is expected to be somewhat decreased owing to 
the effect of quasi-particle damping and mass renormalization. 
For a more quantitative analysis, it is necessary 
to include the normal self-energy corrections. 
In addition, it may be necessary to use more realistic 
three-dimensional electronic structures, 
since the observed resistivity does not show such strong 
two-dimensionality~\cite{ref:WangXF2008, ref:WuG2008} 
as in other unconventional superconductors, 
such as cuprates, Sr$_2$RuO$_4$, and organics. 
On the other hand, the eigenvalues calculated for the 2-band model 
appears to be too small to explain the high $T_{\rm c}$. 
The 2-band model possibly lacks some essential factors, 
e.g., the $xy$ states (equivalently the $X^2-Y^2$ states). 
In fact, as we have seen in the calculations using the 5-band model, 
the superconducting order parameter takes the maximum value 
on the $X^2-Y^2$ state, and the inclusion of at least this orbital 
appears to be necessary to describe iron-pnictide superconductivity. 

We discuss the momentum dependences 
of the anomalous Green's functions. 
As we have seen in \S\ref{sc:Favorable pairing}, 
the anomalous function changes its sign 
between the hole and electron pockets. 
Note that this is a natural result 
of electron correlations. 
In strongly correlated electron systems, 
the motion of electrons is generally affected strongly 
by the Coulomb repulsion. 
Electrons move in metals, avoiding each other, in order 
to reduce the expectation value of the Coulomb interaction. 
The most significant part of the local Coulomb interaction 
in actual solids is the on-site intraorbital interaction, 
i.e., the first term of the right-hand side of eq.~(\ref{eq:H'}): 
\begin{eqnarray}
H'_U &=& \frac{U}{2}\sum_{i}\sum_{\ell}\sum_{\sigma \neq \sigma'}
c_{i\ell\sigma}^{\dag} c_{i\ell\sigma'}^{\dag} 
c_{i\ell\sigma'} c_{i\ell\sigma}. 
\end{eqnarray}
In superconductors, the contribution 
of the superconducting condensate to the expectation value 
of $H'_U$ is evaluated within a simple mean-field approximation: 
\begin{eqnarray}
\delta E_U &=& \langle H'_U \rangle \\
&=& \frac{U}{2}\sum_{i} \sum_{\ell} \sum_{\sigma\neq\sigma'}
\langle c_{i\ell\sigma}^{\dag} c_{i\ell\sigma'}^{\dag} 
c_{i\ell\sigma'} c_{i\ell\sigma} \rangle \\
& \approx & \frac{U}{2N}\sum_{{\mib k}{\mib k}'} \sum_{\ell} \sum_{\sigma} 
F_{\ell\sigma\bar{\sigma}}^{\dag}({\mib k}) F_{\ell\bar{\sigma}\sigma}({\mib k}'), 
\end{eqnarray}
where 
\begin{eqnarray}
F_{\ell\bar{\sigma}\sigma}({\mib k}) &=& \langle c_{{\mib k}\ell\bar{\sigma}} 
c_{-{\mib k}\ell\sigma} \rangle \\
F_{\ell\sigma\bar{\sigma}}^{\dag}({\mib k}) &=& \langle c_{-{\mib k}\ell\sigma}^{\dag} 
c_{{\mib k}\ell\bar{\sigma}}^{\dag} \rangle, 
\end{eqnarray}
and $c_{{\mib k}\ell\sigma}$ is the Fourier transformation of $c_{i\ell\sigma}$. 
In odd-parity spin-triplet pairing states, 
$\delta E_U$ vanishes completely. 
We now consider even-parity spin-singlet pairing states. 
The spin part is decoupled by 
$F_{\ell\sigma\sigma'}({\mib k}) = F_{\ell}({\mib k})
[{\rm i}\sigma_y]_{\sigma\sigma'}$ and 
$F_{\ell\sigma\sigma'}^{\dag}({\mib k}) = F_{\ell}^*({\mib k})
[-{\rm i}\sigma_y]_{\sigma\sigma'}$, and the expectation value 
of the energy is reduced to 
\begin{equation}
\delta E_U = \frac{U}{N} \sum_{\ell} 
\Bigl| \sum_{\mib k} F_{\ell}({\mib k}) \Bigr|^2. 
\end{equation}
It will be easy to see that, in order to reduce $\delta E_U$, 
$|\sum_{\mib k} F_{\ell}({\mib k})|$ should be small for each $\ell$. 
Here note that $F_{\ell}({\mib k})$ is calculated 
by the anomalous Green's functions~\cite{ref:MineevVP1999}: 
\begin{equation}
F_{\ell}({\mib k}) = T \sum_{\omega_n} F_{\ell\ell}(k). 
\end{equation}
The momentum dependence and nodal positions 
of $F_{\ell\ell}(k)$ (except absolute magnitudes) 
are generally almost the same at all frequencies. 
Therefore, we may expect that the momentum dependence 
of $F_{\ell}({\mib k})$ is approximately 
the same as that of $F_{\ell\ell}(k)$. 
As we have seen in \S\ref{sc:Favorable pairing}, 
the maximum and minimum of $F_{\ell\ell}(k)$ 
in momentum space have almost the same absolute magnitude 
but opposite signs to each other, 
and consequently $\sum_{\mib k} F_{\ell\ell}(k)$ 
is suppressed to a small value for each $\ell$. 
Thus, we find that the calculated momentum dependences 
of the anomalous functions $F_{\ell\ell}(k)$ 
are indeed such that the on-site intraorbital 
correlation energy $\delta E_U$ is reduced. 
Using eqs.~(\ref{eq:FF}) and (\ref{eq:UU}), 
we obtain the band representation 
\begin{equation}
\biggl|\sum_{\mib k} F_{\ell} ({\mib k}) \biggr| = 
\biggl|\sum_{\mib k} \sum_a |U_{\ell a}({\mib k})|^2 F_a ({\mib k}) \biggr|. 
\end{equation}
In many other unconventional superconductors 
including $d_{x^2-y^2}$-wave superconductors 
such as some heavy-fermion superconductors, 
this value will be suppressed by the sign change 
of $F_a ({\mib k})$ on each band $a$. 
On the other hand, in iron-pnictide superconductors, 
it is suppressed by the sign change of $F_a ({\mib k})$ 
not on each band but among different bands. 
Thus, we naturally expect that the superconducting order parameter 
should always take nearly the same absolute value 
but opposite signs between the hole pockets 
around the $\Gamma$ point and the electron pockets around the M point, 
in order to reduce the correlation energy $\delta E_U$. 
The superconducting gap function will also always take nearly 
the same absolute values between the hole pockets 
and the electron pockets, although the gap magnitude 
itself will depend on the temperature and doping level. 
This is expected to be confirmed by experiments, such as 
comprehensive ARPES studies on the band dependence 
of gap magnitude at various temperatures and doping levels 
in various iron-pnictide superconductors. 
If experiments indeed confirm this, they will support the realization 
of the $s_\pm$-wave pairing symmetry in iron-pnictide superconductors. 

It will be interesting to compare our results with those of other 
theoretical studies. 
The $s_\pm$-wave or extended $s$-wave pairing states 
have been obtained by many other theoretical studies. 
Among them, the RPA calculations were performed by Kuroki 
et al. for the same 5-band Hubbard model~\cite{ref:KurokiK2008} 
(The 5-band model used in our present study was constructed by them), 
and by Yanagi et al. for a more complex $d$-$p$ model 
including the As4$p$ orbitals explicitly~\cite{ref:YanagiY2008}. 
We speculate that the $s_\pm$-wave pairing state will 
be more robustly favorable against the next most favorable pairing state, 
i.e., the $d_{x^2-y^2}$-wave state, in the present perturbative 
study than in the RPA studies. 
For a higher doping level $x=0.3$ ($n=6.3$), 
where the hole pocket $\alpha$' 
around $(\pi, \pi)$ does not exist, 
the $d_{x^2-y^2}$-wave state might become the most favorable 
as mentioned in ref.~\ref{ref:KurokiK2008}, but the eigenvalues will not 
be large enough to explain the high $T_{\rm c}$. 
Our perturbative results suggest that the superconducting order parameter 
(and superconducting gap) does not have any nodes on the Fermi surface 
in the whole doping region, while nodes appear on the electron 
pockets around the M point in the RPA results of ref.~\ref{ref:KurokiK2008}, 
despite the fact that we have used the same model as Kuroki et al. used. 
The origin of this discrepancy is unclear. 
Other more recent RPA results by Ikeda suggest that there 
are not nodes on the Fermi surface~\cite{ref:IkedaH2008}. 
Yanagi et al. suggested that the $s_\pm$-wave state 
is the most favorable in a $d$-$p$ model with reasonable parameters, 
but the $d_{x^2-y^2}$-wave pairing state (the $d_{XY}$-wave state 
in the original folded Brillouin zone) becomes the most favorable, 
overcoming the $s_\pm$-wave state for small Hund's coupling 
or $J=0$~\cite{ref:YanagiY2008}. 
On the other hand, in our present perturbative results 
using the Hubbard model, we could not find any sets of Coulomb 
integrals for which the $d_{x^2-y^2}$-wave pairing state 
becomes the most favorable. 
We speculate that this difference originates 
from the difference in electronic structures used 
for the calculations (i.e., the 5-band Hubbard model or 
the more complex $d$-$p$ model), 
rather than that in approximation methods 
(i.e., third-order perturbation theory or RPA). 
In fact, according to recent RPA results for the 5-band Hubbard 
model~\cite{ref:IkedaH2008}, the $d_{x^2-y^2}$-wave pairing state 
does not overcome the $s_\pm$-wave state even for $J=0$, 
as obtained from our present third-order perturbation theory. 
Concerning the superconducting mechanism, we consider that 
the magnetic fluctuations due to the Fermi surface nesting 
between the hole and electron pockets are an important 
key factor, as considered in many other theoretical 
studies~\cite{ref:MazinII2008, ref:KurokiK2008, 
ref:YanagiY2008, ref:WangF2008, ref:ChubukovAV2008, ref:IkedaH2008}. 

We can now understand why the transition temperature drops suddenly
around the high-doping regions where the hole pocket 
around $(\pi, \pi)$ or the electron pocket 
around the M point disappears. 
This can be regarded as a natural result of the $s_\pm$-wave 
pairing due to electron correlations. 
For the higher doping levels, the Fermi surface nesting 
between the hole pockets and electron pockets cannot occur, 
because some of the nesting Fermi surfaces disappear. 
In addition, the sign change of the order parameter between them 
can no longer occur, and enhancement of the on-site correlation 
energy $\delta E_U$ will inevitably destabilize 
the $s_\pm$-wave pairing state. 
We expect that the disappearance of superconductivity 
at the doping levels where the hole pockets 
or electron pockets disappear will be observed 
by systematic experiments in which 
both the Fermi surface and transition temperatures 
are measured for various doping levels. 

Finally, we mention that the 5-band model used 
in the present study is unsatisfactory regarding the following point. 
The volumes of the Fermi pockets appear to be quantitatively different 
from those of actual Fermi surfaces observed in ARPES experiments: 
the hole pockets have smaller volumes, 
while the electron pockets have larger volumes, 
than those in the ARPES results. 
This is the reason why the calculated transition temperature 
drops suddenly at a relatively low electron-doping level (about $x=0.18$) 
and survives even up to a high hole-doping level (about $x=-0.3$), 
which is quantitatively inconsistent with experimental results. 
In actual systems, the superconductivity survives 
above the electron-doping level of $x=0.18$, and vanishes 
at the high hole-doping level of $x=-0.3$. 
Thus, models possessing precise electronic structures 
and fermiology that are completely consistent with actual systems 
will be desirable for more quantitative studies. 

\section{Conclusions}

In the present work, we have studied the high-$T_{\rm c}$ 
superconductivity in iron pnictides by using multi-band Hubbard 
models for iron 3d electrons and by solving the Eliashberg equation 
within the third-order perturbation theory. 
The unconventional $s_\pm$-wave pairing symmetry is the most favorable 
and is highly likely to be realized, since it explains 
the high transition temperatures of iron pnictides. 
The superconducting gap is not expected to have any nodes 
on the Fermi surfaces in the whole doping region. 
This unconventional pairing symmetry is consistent 
with the results of many other theoretical and experimental studies. 
The microscopic origin of this unconventional pairing 
is the local Coulomb interaction among the iron 3d electrons. 
Thus, we can regard this high-$T_{\rm c}$ 
superconductivity of iron pnictides 
as a natural result of the electron correlation, 
as in the cases of other established 
correlation-induced (non-phonon-mediated) 
superconductors, such as cuprate, ruthenate, organic, 
and some heavy-fermion superconductors. 
An important difference from other unconventional 
superconductors is that the order parameter changes 
its sign not on the Fermi surface but between the Fermi pockets. 
We hope that further comprehensive measurements 
such as those mentioned in \S\ref{sc:Discussion} will 
be carried out in order to confirm the realization 
of this intriguing pairing symmetry 
in iron-pnictide superconductors. 

\section*{Acknowledgements}
The author would like to thank 
Prof. Yasumasa Hasegawa, Dr. Hiroaki Ikeda, 
Prof. Hiroshi Kontani, Prof. Kazuhiko Kuroki, 
Dr. Kazuma Nakamura, Dr. Seiichiro Onari, 
Prof. Takasada Shibauchi, and Prof. Kosaku Yamada
for valuable communications. 
Numerical work was carried out at the Yukawa Institute 
Computer Facility of Kyoto University. 

\bibliography{99}

\end{document}